\documentclass[12pt]{article}
\usepackage{amsmath,amssymb,amsthm,mathtools,booktabs,array,lmodern,float}
\usepackage[paperwidth=210mm,paperheight=248mm,left=20mm,right=6mm,top=5mm,bottom=5mm,includeheadfoot,headheight=16pt,headsep=3mm,footskip=8mm]{geometry}
\usepackage{fancyhdr}
\usepackage{xurl}
\usepackage{hyperref}\hypersetup{hidelinks,pdftitle={Increasing Quantum LDPC Code Distance through Explicit Extensions},pdfauthor={Kenta Kasai}}
\usepackage{needspace}
\usepackage{graphicx}
\usepackage{enumitem}\setlist{nosep,leftmargin=2em}
\theoremstyle{definition}

\renewcommand{\proofname}{Proof}

\newcommand{\F}{\mathbb F_2}
\newcommand{\Z}{\mathbb Z}

\newcommand{\rank}{\operatorname{rank}}

\newcommand{\diag}{\operatorname{diag}}

\newcommand{\code}[3]{[\![#1,#2,#3]\!]}

\title{Increasing Quantum LDPC Code Distance\\through Explicit Extensions}
\author{Kenta Kasai\\\normalsize Institute of Science Tokyo}
\date{November 21, 2026}
\begin{document}
\maketitle
\thispagestyle{fancy}
\begin{abstract}
We give an explicit construction of quantum low-density parity-check (LDPC) codes without searching for candidates at each size. A fixed pair partition (PP) and fixed integer parameters determine the parity-check matrices at a specified size. Using affine permutation matrices (APMs), we construct a $\code{3072}{772}{32}$ code with column weight~3 and check weight~8 in each of the X and Z parity-check matrices. A construction based on circulant permutation matrices (CPMs) gives $\code{1270}{260}{\ge34}$, with column weight~4 and check weight~10. We prove that the construction yields commuting checks and verify the resulting distances using exhaustive search and logical operators. For the main construction, we conjecture the existence of a finite structural distance upper bound and distance improvement toward that bound as the size increases.
\end{abstract}

\section{Background and objective}
Quantum low-density parity-check (LDPC) codes protect quantum information while limiting both the number of qubits in each check and the number of checks involving each qubit~\cite{MMM}. In the Calderbank--Shor--Steane (CSS) construction, orthogonality of the X and Z parity-check matrices must be reconciled with the absence of low-weight nontrivial logical operators~\cite{CSS,Steane}. Hypergraph-product constructions achieve a positive rate and distance proportional to the square root of the blocklength~\cite{TZ}. Subsequent constructions give quantum LDPC families whose rate and relative minimum distance are both bounded below by positive constants~\cite{PKGood}.

At finite blocklengths, the weights of the checks actually used matter alongside blocklength, dimension, and minimum distance. Studies of generalized bicycle codes~\cite{PKFinite} and low-overhead quantum memory based on bivariate bicycle codes~\cite{BBMemory} have advanced the concrete design of sparse checks with algebraic structure. We study the extension of an already good code to another size while keeping its column and check weights fixed. This requires a rule that preserves orthogonality across sizes and a means of verifying the distances of the resulting codes.

Constructions using circulant permutation matrices (CPMs) and affine permutation matrices (APMs) provide a suitable starting point. Affine-permutation constructions control commutativity among permutations to satisfy orthogonality~\cite{Barrier}. Pair-partition (PP) constructions guarantee orthogonality by pairing block columns so that the two contributions in each pair cancel~\cite{PP,DP}. Relations among column weight, check weight, distance, and blocklength have been established, and searches that screen pair partitions and exponent arrays have produced good finite-length codes~\cite[Sec.~III.2]{DP}. This leads to the question addressed here: can the construction information in a good original code be reused to improve distance without searching for candidates at each size?

Increasing the size also requires attention to distance restrictions imposed by the construction. Classical quasi-cyclic LDPC codes have minimum-distance upper bounds determined by their block structure~\cite{SV}. For classical CPM--LDPC codes, an explicit construction fixes a single integer exponent array and attains a structural distance upper bound for all sufficiently large $P$~\cite[Theorem~3]{CPMAttain}. Quantum codes additionally require orthogonality and a distinction between low-weight stabilizers and nontrivial logical operators. The central tasks are therefore to identify the algebraic conditions inherited from the original code and to verify the quantum distance after extension.

The behavior of random constructions also depends on the sampling distribution. For full $J\times L$ arrays of arbitrary permutation matrices sampled uniformly subject to CSS orthogonality, the relative minimum distance converges to zero in probability as $P\to\infty$, for fixed $J\ge3$ and even $L>2J$~\cite[Theorem~4.1]{RandomCSS}. This asymptotic result for a specified distribution does not determine the finite-length success rate of searches restricted to CPMs or APMs. Here we investigate a construction whose rule is fixed from a good original code.

Distance computation includes both searches for low-weight codewords that give upper bounds and methods that certify the absence of lighter codewords. For classical linear codes, representative approaches are probabilistic searches for low-weight codewords~\cite{Leon,Stern} and the Brouwer--Zimmermann algorithm, which enumerates codewords using multiple information sets until upper and lower bounds meet~\cite{GrasslBZ,LT}. Improvements include the choice of information sets~\cite{LT} and parallel implementations on distributed-memory machines~\cite{ParallelBZ}. These methods support the complementary tasks of finding promising candidates quickly and certifying their distances.

For quantum codes, stabilizers must be excluded so that the search minimizes the weight of a nontrivial logical operator. Prior work includes upper-bound searches using random information sets~\cite{QDistRnd}, fast Brouwer--Zimmermann implementations for symplectic weight~\cite{QuantumBZ}, and irreducible-cluster searches exploiting sparse checks~\cite{DKP}. Recent benchmarks also compare exact methods based on satisfiability and mixed integer programming~\cite{DistanceBench}. For PP codes, permutation symmetries and weight lower bounds reduce the exhaustive search space~\cite{PP,DP,QEDist}. For our constructions, exhaustive support search with symmetry reduction and pruning certifies lower bounds, which we compare with upper bounds from verified nontrivial logical operators.

Our contribution is a common procedure that extracts a pair partition and integer parameters from a general CPM/APM--PP code and uses them to generate parity-check matrices at another size. Once the generation rule is fixed, it gives a code at each admissible size without distance-based candidate screening; the construction is explicit in this sense. Section~\ref{sec:construction} proves conditions that preserve orthogonality and column and check weights. Section~\ref{sec:instances} presents the codes generated by fixed rules and the verified distance improvements. 

\section{Explicit construction with fixed parameters}\label{sec:construction}
The aim is to extract a rule that preserves orthogonality from a good original code and seek distance improvement without searching for candidates at each size. A low-weight logical operator selects a small set of columns whose contributions cancel at every relevant check. The index congruences that permit its support depend on the size, so changing the size may invalidate the same support arrangement. Meanwhile, orthogonality between the X and Z checks must be retained. We therefore keep the pair partition that guarantees orthogonality and generate the slopes and shifts of the permutations for the new size.

This construction builds on exponent-array generation from solution spaces of the PP conditions~\cite[Sec.~IV-B]{PP}\cite[Sec.~III.2]{DP} and commuting-check constructions using APMs~\cite{Barrier}. Smith normal form~\cite[pp.~314--318]{Smith} is also used to solve shift congruences in the public APM generator~\cite{Seed}. We use these tools to extract integer parameters reproducing the arrays of a good original code and reuse both the parameters and the generation rule at another size. This preserves orthogonality and column and check weights while allowing the conditions for low-weight logical operators to change. New low-weight logical operators may also appear, so distance improvement is verified after construction. The procedure below applies this idea to general CPM/APM--PP inputs and establishes its validity.

\subsection{Constructing a new code from an original code}\label{sec:fixeddata}
Let $P_0$ be the original size and $\boldsymbol{a}_0,\boldsymbol{b}_0$ the original slope and shift arrays. We first use the pair partition to specify orthogonality conditions at every size. We then fix parameters representing the original code among the arrays satisfying those conditions and use them to construct arrays at a specified size $P$.
Let $P\ge2$ and $\Z_P=\Z/P\Z$. For $a\in\Z_P^\times$ and $b\in\Z_P$, define $A(a,b)_{v,u}=\mathbf1\{v=au+b\pmod P\}$. Here $a$ is the slope and $b$ is the shift; a CPM is the case $a=1$. For a positive even integer $L$ and $J\ge1$, set
\begin{equation}\label{eq:checks}
 H_X=\bigl(A(a^X_{i\ell},b^X_{i\ell})\bigr),\qquad
 H_Z=\bigl(A(a^Z_{j\ell},b^Z_{j\ell})\bigr),
 \quad 0\le i,j<J,\quad0\le\ell<L.
\end{equation}
The orthogonality condition for a CSS code~\cite{CSS,Steane} is $H_XH_Z^{\mathsf T}=0$, and its dimension is $k=LP-\rank H_X-\rank H_Z$. Ranks and row spaces are taken over $\F$. The distance $d$ is the minimum weight of a logical Pauli operator outside the stabilizer.

To preserve orthogonality, we retain the original pair partition $M_{ij}$ of column indices for each pair of block rows $(i,j)$. At the new size, the two contributions within each of these same pairs must still cancel. The PP conditions to be satisfied at both sizes are, for every $\{u,v\}\in M_{ij}$,
\begin{equation}\label{eq:generalpair}
 \rho_{iju}=\rho_{ijv},\qquad
 b^Z_{ju}-\rho_{iju}b^X_{iu}=b^Z_{jv}-\rho_{ijv}b^X_{iv},
 \qquad \rho_{ij\ell}=a^Z_{j\ell}(a^X_{i\ell})^{-1}.
\end{equation}
These conditions make the permutation products equal within each pair, implying orthogonality~\cite[Theorem~4]{PP}. We order the $N=2JL$ components with the X array first and the Z array second, using row-major order within each array. The order and orientation of pairs and the row order are also fixed.

We first solve the slope conditions. Simply using the original values with a new modulus need not satisfy these conditions, so we parametrize all admissible slopes. Write the slope conditions as $\prod_{q=0}^{N-1}a_q^{T_{rq}}=1$. Each row of the integer matrix $T$ places $(-1,1,1,-1)$ at positions $(Xiu,Xiv,Zju,Zjv)$. Let $UTQ=\diag(e_0,\ldots,e_{N-1})$ be a Smith normal form~\cite[pp.~314--318]{Smith}, padding missing diagonal entries with zeros. The matrices $U,Q$ are invertible over the integers. Since the unit group is abelian, all slope solutions have the form
\begin{equation}\label{eq:general-slopes}
 a_q(P)=\prod_{j=0}^{N-1}r_j(P)^{Q_{qj}}\pmod P,
 \qquad r_j(P)\in\Z_P^\times,\quad r_j(P)^{e_j}=1.
\end{equation}
For $e_j=0$, the last condition holds automatically. Negative exponents use multiplicative inverses of units.

After choosing the slopes, we express the original shifts through fixed parameters. At each size, the admissible shifts solve the following linear congruences:
\begin{equation}\label{eq:shift}
 B_P\boldsymbol{b}=\boldsymbol{0}\pmod P,\qquad
 (B_P)_{r,(Xiu,Xiv,Zju,Zjv)}=(-\rho_{iju},\rho_{iju},1,-1).
\end{equation}
We represent $\rho$ by an integer in $0,\ldots,P-1$. Compute $U_PB_PV_P=\diag(d_{P,0},\ldots)$ using the same Smith-normal-form convention and pad zero diagonal entries. All shift solutions are then given by
\begin{equation}\label{eq:fixedtheta}
 \boldsymbol{b}(P)=G(P)\boldsymbol{\theta}\pmod P,\qquad
 G(P)=V_PC_P,\quad C_P=\diag\!\left(\frac{P}{\gcd(P,d_{P,j})}\right),
 \quad\boldsymbol{\theta}\in\Z^N,
\end{equation}
where $\gcd(P,0)=P$. Since the basis is not unique, the procedure used to compute the Smith normal form is fixed as well. At the original size, we extract an integer vector $\boldsymbol{\theta}$ satisfying $\boldsymbol{b}_0=G(P_0)\boldsymbol{\theta}\pmod{P_0}$ and use the same vector at the new size. The vector $\boldsymbol{\theta}$ gives the coordinates of the original shifts in this parametrization. Recomputing $G(P)$ at a new size retains these coordinates while producing shifts that satisfy the new orthogonality conditions.

Algorithm~1 extracts data from the original code once, and Algorithm~2 constructs the matrices at each specified size. The first checks reproduction of the original arrays; the second computes directly from the fixed data, without evaluating distances to select a candidate.

\medskip\noindent\textbf{Algorithm 1: Extract integer parameters from the original code.}\par
\textbf{Input:} $J,L,P_0$ and the original slope and shift arrays. A pair partition $M$ may also be supplied.\par
\textbf{Output:} $M,Q,(e_j),\boldsymbol{\lambda},\boldsymbol{\theta}$ and fixed computational conventions.
\begin{enumerate}[label=\arabic*.,leftmargin=1.8em]
\item Check that all slopes are units and that \eqref{eq:generalpair} holds. If $M$ is not supplied, group columns with the same $(\rho_{ij\ell},b^Z_{j\ell}-\rho_{ij\ell}b^X_{i\ell})$ for each $(i,j)$, and pair consecutive indices within each group. If a group has odd cardinality, stop because the input does not satisfy the PP conditions. Also stop if a supplied $M$ fails the conditions.
\item Compute a Smith normal form of $T$ and set $r_{0,j}=\prod_q a_{0,q}^{(Q^{-1})_{jq}}\pmod{P_0}$. Order $\mathcal R_e(P)=\{r\in\{1,\ldots,P-1\}:\gcd(r,P)=1,\ r^e\equiv1\pmod P\}$ increasingly. Store the position of $r_{0,j}$ in $\mathcal R_{e_j}(P_0)$ as a zero-based integer $\lambda_j$.
\item Compute $B_{P_0},V_{P_0},C_{P_0}$. Represent each component of $\boldsymbol{y}_0=V_{P_0}^{-1}\boldsymbol{b}_0\pmod{P_0}$ in $0,\ldots,P_0-1$, and set $\theta_j=y_{0,j}/(C_{P_0})_{jj}$. Each quotient is an integer because the original shifts satisfy \eqref{eq:shift}.
\item Check that \eqref{eq:general-slopes} and \eqref{eq:fixedtheta} reproduce the original arrays, and save the fixed output.
\end{enumerate}

\medskip\noindent\textbf{Algorithm 2: Construct a code of a specified size from fixed parameters.}\par
\textbf{Input:} The output of Algorithm~1 and an integer $P\ge2$.\par
\textbf{Output:} Slope and shift arrays and the matrices $H_X,H_Z$.
\begin{enumerate}[label=\arabic*.,leftmargin=1.8em]
\item For each $j$, choose $r_j(P)$ at position $\lambda_j\bmod|\mathcal R_{e_j}(P)|$ in $\mathcal R_{e_j}(P)$, and compute the slopes using \eqref{eq:general-slopes}. Sets with the same $e_j$ need only be computed once.
\item Form $B_P$ from these slopes, and use its Smith normal form to compute $G(P)=V_PC_P$.
\item Use the fixed $\boldsymbol{\theta}$ in \eqref{eq:fixedtheta} to compute the shifts. Reshape them into arrays and assemble the parity-check matrices in \eqref{eq:checks}.
\end{enumerate}
The set $\mathcal R_e(P)$ always contains~1, so this procedure is defined for every $P\ge2$. It can be computed in finite time by checking greatest common divisors and modular powers for $1,\ldots,P-1$, without screening candidate codes by distance. For a CPM input, all $r_{0,j}=1$ and $\lambda_j=0$, so the output is also CPM-based. This default rule handles general inputs; it does not guarantee growth in distance or dimension, or efficient computation at large sizes.

For a valid CPM/APM--PP input, Algorithms~1 and~2 reproduce the original arrays at $P=P_0$ and give orthogonal parity-check matrices for every $P\ge2$. The blocklength is $LP$, the column weight of each parity-check matrix is $J$, and the check weight is $L$. Every vector in either kernel has even weight. Positive quantum dimension must be checked separately.

These properties follow from the construction of the slopes and shifts. The matrices $U,Q$, which are invertible over the integers, give invertible transformations of multiplicative row relations and variables over any abelian group. Thus the slope conditions are equivalent to $r_j^{e_j}=1$. Algorithm~1 stores the positions of the original $r_{0,j}$ and satisfies $C_{P_0}\boldsymbol{\theta}\equiv V_{P_0}^{-1}\boldsymbol{b}_0$, reproducing the original arrays. At a new size, the slope conditions and $B_PG(P)=0\pmod P$ still hold, so the PP conditions imply orthogonality. Each block is a permutation matrix, which fixes the row and column weights. The sum of all rows in any one block row is the all-one vector, implying that every kernel vector has even weight~\cite[Sec.~V-A]{PP}.

\subsection{Choosing the generation rule}
A generation rule reproducing a given original code is not unique. The fixed data therefore include not only $\boldsymbol{\theta}$ but also the rules for computing the slopes and $G(P)$ at each size. The general procedure also permits other fixed rules for $r_j(P)\in\mathcal R_{e_j}(P)$ that reproduce the original slopes. The shifts may likewise use a matrix $G(P)$ with fewer columns, provided that $B_PG(P)=0\pmod P$ and reproduction of the original shifts are verified. In that case, solve $G(P_0)\boldsymbol{\theta}\equiv\boldsymbol{b}_0$ and fix integer representatives. If no solution exists, use the default parametrization of all solutions. Any coordinate normalization must be checked for permutation equivalence to the original code. These choices are fixed before distance verification at each size, and different choices define different families. The conditions above establish orthogonality and weight preservation; which rules improve distance is determined by verifying the resulting codes.

\section{Code constructions and distance verification}\label{sec:instances}\label{sec:specializations}
Applying the preceding procedure to concrete codes gives the fixed generation rules and verified distance improvements below.

The examples use a publicly available APM--PP code obtained by search~\cite{Seed} and previously reported CPM--PP codes. Table~\ref{tab:codes} lists the original code in each family and its extensions under the same rule.

The relative distance $d/n$ decreases for the main APM example. For CPM--PP codes with $J\ge2$, odd size $P$, $L\ge2J+1$, and rank $J(P-1)+1$ for both parity-check matrices, the distance is at most $(J+1)!$~\cite[Sec.~V-B, Theorem~3]{PP}~\cite[App.~A]{DP}. This bound, based on the permanent construction for classical quasi-cyclic codes~\cite{SV}, is~24 for $(J,L)=(3,8)$. Its hypotheses do not apply to the main construction.

\begin{table}[H]\centering\small
\caption{Codes obtained without candidate screening at each size. Here $n$ is the blocklength, $k$ the dimension, $d$ the minimum distance, $J$ the column weight of both the X and Z parity-check matrices, $L$ the check weight (row weight), and $g$ the common girth of the X and Z Tanner graphs. The original APM--PP distance~26 is established here; the original distance~22 for $L=10$ is a previously reported value; $\ge$ denotes a verified distance lower bound.}\label{tab:codes}
\begin{tabular}{lrcrrr}\toprule
Construction & $P$ & $[[n,k,d]]$ & $J$ & $L$ & $g$\\\midrule
APM--PP (original code)~\cite{Seed} & 192 & $\code{1536}{388}{26}$ & 3 & 8 & 8\\
APM--PP (extension) & 384 & $\code{3072}{772}{32}$ & 3 & 8 & 8\\
\midrule
CPM--PP (original code)~\cite[Table~8]{DP} & 37 & $\code{370}{80}{22}$ & 4 & 10 & 6\\
CPM--PP (extension) & 73 & $\code{730}{152}{24}$ & 4 & 10 & 6\\
CPM--PP (extension) & 127 & $\code{1270}{260}{\ge34}$ & 4 & 10 & 6\\
\midrule
CPM--PP (original code)~\cite[Table~8]{DP} & 23 & $\code{276}{98}{14}$ & 4 & 12 & 6\\
CPM--PP (extension) & 37 & $\code{444}{154}{16}$ & 4 & 12 & 4\\
CPM--PP (extension) & 53 & $\code{636}{218}{20}$ & 4 & 12 & 6\\
\bottomrule\end{tabular}
\end{table}

For the first APM family and the CPM family with check weight~10, we fix the generation rules described below.

\needspace{10\baselineskip}
For the main APM example, fix integer matrices $S^X,S^Z$ satisfying
\begin{equation}\label{eq:integerS}
 S^X_{iu}-S^X_{iv}=S^Z_{ju}-S^Z_{jv}\qquad(\{u,v\}\in M_{ij}),
\end{equation}
and set
\begin{equation}\label{eq:slope}
 a^{X,Z}_{i\ell}(P)=1+DS^{X,Z}_{i\ell}\pmod P,\qquad D^2\equiv0\pmod P.
\end{equation}
The inverse is $1-DS$, so \eqref{eq:integerS} makes the relative slopes equal within each pair. This is therefore a valid choice in \eqref{eq:general-slopes}. For the original code with $P_0=192,D_0=96$, compute $s_q=(a_{0,q}-1)/96$, take the reduced row echelon form of \eqref{eq:integerS} over the rationals, and set its free components to the corresponding $s_q$. Solve for the remaining components, verify integrality and reproduction of the original slopes, and fix $S$.

This example uses $G(P)=V_PC_P$. The shift-extraction step of Algorithm~1 fixes a 48-component vector $\boldsymbol{\theta}$~\cite{CodeData}. The sizes follow
\begin{equation}\label{eq:family}
 P_t=192\cdot2^t,\qquad D_t=96\cdot2^{\lfloor t/2\rfloor},\qquad
 h_t=P_t/D_t=2^{1+\lceil t/2\rceil}\quad(t\ge0).
\end{equation}
Since $D_t^2/P_t$ is either~48 or~24, every size gives orthogonal parity-check matrices with the original column and check weights.

For the CPM family with check weight~10, all slopes are~1. Normalize the original exponent arrays by subtracting the first X row from every row, then subtracting each row's first entry from that row. Add the resulting zero-entry conditions to the integer PP equations. Clear denominators in a rational nullspace basis to fix $G=(\boldsymbol{k}_1,\ldots,\boldsymbol{k}_r)$, and solve the congruence described above. If this representation does not include the original shifts, revert to the default parametrization. In the displayed example, $r=3$. The coefficients $(17,31,33)$ obtained modulo~37 become $\boldsymbol{\theta}=(1,4,15)^{\mathsf T}$ after global normalization by the inverse of~17. Permutation equivalence under this normalization was also verified. The fixed $G$ and $\boldsymbol{\theta}$ generate the shifts at each size.

For check weight~12, we apply Algorithms~1 and~2 directly to the original code in Table~\ref{tab:codes}. Distance is not monotone in size: the same fixed rule with check weight~12 gives distances~16 and~14 at $P=43$ and~47, respectively~\cite{CodeData}.

Distance verification uses exhaustive search that incrementally extends supports~\cite{DKP}~\cite[Sec.~V-A]{PP}~\cite[App.~B]{DP}. It combines search-space reduction using verified permutation symmetries~\cite{PP,DP}, pruning by lower bounds on the additional weight required for completion~\cite{PP,DP,QEDist}, and rejection of candidates whose weight decreases after adding a stabilizer~\cite{DP,QEDist}. Completion of the entire partitioned search space is audited.

Additional acceleration could use known weight reductions from sums of two generators~\cite[App.~D.2]{DP} to reject partial supports. Each qubit also participates in exactly one check in each block row. Two block rows therefore define a graph whose vertices are their checks and whose edges are qubits. Finding a minimum edge set whose odd-degree vertices are the unsatisfied checks---a minimum T-join~\cite{EJ}---could strengthen the lower bound on additional weight. The speedups from these additions have not been measured, and they were not used to establish the displayed distance guarantees.

We checked that matrices reconstructed from the fixed data match the search inputs. Exhaustive search using the existing methods above, together with an audit of the partitioned search space, excluded low-weight nontrivial logical operators. We based the computations on existing distance-verification methods and software~\cite{PP,DP,QEDist}, adding partitioned parallel execution and result auditing. For some computations, we also added checkpoint and resume support while preserving the pruning rules and search order. The versions, inputs, and search ranges were recorded. Both logical types were covered either by searches on both sides or by a verified X/Z exchange symmetry. The even-weight property of the kernels established in Section~\ref{sec:fixeddata} then gives the lower bounds in the table. For exact distances, a nontrivial logical operator of the matching weight was verified in physical coordinates. For the original code with check weight~10, the distance is inherited after verifying permutation equivalence to the previously reported code~\cite[Table~8]{DP}.

Construction data, matrices, logical operators, and audit summaries are publicly available~\cite{CodeData}. The cited data release gives a lower bound of~32 at $P=127$; the supplementary material includes the additional certificates and independent audit supporting the lower bound of~34 reported here. A nontrivial logical operator of weight~70 has also been verified, giving the rigorous interval $34\le d\le70$. Full exhaustive-search records and audit procedures for codes with established exact distances are published as release attachments with checksums~\cite{CodeEvidence}. Independently reproducing an exclusion requires rerunning the saved implementation on the recorded inputs and search ranges; an audit summary alone does not reproduce the search.

\section{Conclusion and distance improvement toward a structural bound}\label{sec:conjecture}
The verified distance improvements motivate a conjecture about further improvement toward a structural upper bound and identify the remaining questions.

A fixed rule removes the need for candidate screening at each size and yields distance improvement at fixed check weight. The construction can change the index-consistency conditions for low-weight operators while retaining orthogonality through the PP conditions. For the fixed data of the main APM construction and the family in \eqref{eq:family}, we conjecture a finite, size-independent structural distance upper bound and improvement toward that bound as the size increases.

The value and attainability of the bound remain open, and monotone growth is not assumed. Further work should identify which properties of the parameters fixed from the original code lead to improvement and establish distance lower bounds that replace exhaustive search at each size. Improvement at the same length requires changing the generation data.

Further work should also isolate the contribution of each acceleration method by comparing runtimes on the same codes under identical computing conditions.

\needspace{6\baselineskip}


\begin{thebibliography}{10}
\small
\setlength{\itemsep}{0pt}
\bibitem{MMM} D.~J.~C. MacKay, G. Mitchison, and P.~L. McFadden, ``Sparse-graph codes for quantum error correction,'' \emph{IEEE Transactions on Information Theory}, vol.~50, no.~10, pp.~2315--2330, 2004. \url{https://doi.org/10.1109/TIT.2004.834737}.
\bibitem{CSS} A.~R. Calderbank and P.~W. Shor, ``Good quantum error-correcting codes exist,'' \emph{Physical Review A}, vol.~54, pp.~1098--1105, 1996. \url{https://doi.org/10.1103/PhysRevA.54.1098}.
\bibitem{Steane} A.~M. Steane, ``Error correcting codes in quantum theory,'' \emph{Physical Review Letters}, vol.~77, pp.~793--797, 1996. \url{https://doi.org/10.1103/PhysRevLett.77.793}.
\bibitem{TZ} J.-P. Tillich and G. Z\'emor, ``Quantum LDPC codes with positive rate and minimum distance proportional to the square root of the blocklength,'' \emph{IEEE Transactions on Information Theory}, vol.~60, no.~2, pp.~1193--1202, 2014. \url{https://doi.org/10.1109/TIT.2013.2292061}.
\bibitem{PKGood} P. Panteleev and G. Kalachev, ``Asymptotically good quantum and locally testable classical LDPC codes,'' in \emph{Proceedings of the 54th Annual ACM SIGACT Symposium on Theory of Computing (STOC)}, pp.~375--388, 2022. \url{https://doi.org/10.1145/3519935.3520017}.
\bibitem{PKFinite} P. Panteleev and G. Kalachev, ``Degenerate quantum LDPC codes with good finite length performance,'' \emph{Quantum}, vol.~5, art.~585, 2021. \url{https://doi.org/10.22331/q-2021-11-22-585}.
\bibitem{BBMemory} S. Bravyi, A.~W. Cross, J.~M. Gambetta, D. Maslov, P. Rall, and T.~J. Yoder, ``High-threshold and low-overhead fault-tolerant quantum memory,'' \emph{Nature}, vol.~627, pp.~778--782, 2024. \url{https://doi.org/10.1038/s41586-024-07107-7}.
\bibitem{Barrier} K. Kasai, ``Breaking the orthogonality barrier in quantum LDPC codes,'' \emph{Quantum}, vol.~10, art.~2205, 2026. \url{https://doi.org/10.22331/q-2026-09-09-2205}.
\bibitem{PP} K. Okada and K. Kasai, ``Pair-partition constructions for CPM-based quantum LDPC codes,'' \url{https://arxiv.org/abs/2607.14091v4}, 2026.
\bibitem{DP} J.~Y. Lee, K. Okada, N. Maskara, K. Kasai, and H. Zhou, ``Design principles for ultra-high-rate quantum codes,'' \url{https://arxiv.org/abs/2609.30069v1}, 2026.
\bibitem{SV} R. Smarandache and P.~O. Vontobel, ``Quasi-cyclic LDPC codes: Influence of proto- and Tanner-graph structure on minimum Hamming distance upper bounds,'' \emph{IEEE Transactions on Information Theory}, vol.~58, no.~2, pp.~585--607, 2012. \url{https://doi.org/10.1109/TIT.2011.2173244}.
\bibitem{CPMAttain} K. Kasai, ``CPM-LDPC codes attaining the minimum-distance bound,'' \url{https://arxiv.org/abs/2609.16836v1}, 2026.
\bibitem{RandomCSS} K. Kasai, ``Vanishing relative distance of uniformly random CSS LDPC codes built from permutation matrices,'' \url{https://arxiv.org/abs/2611.21389v1}, 2026.
\bibitem{Leon} J.~S. Leon, ``A probabilistic algorithm for computing minimum weights of large error-correcting codes,'' \emph{IEEE Transactions on Information Theory}, vol.~34, no.~5, pp.~1354--1359, 1988. \url{https://doi.org/10.1109/18.21270}.
\bibitem{Stern} J. Stern, ``A method for finding codewords of small weight,'' in \emph{Coding Theory and Applications}, Lecture Notes in Computer Science, vol.~388, pp.~106--113, Springer, 1989. \url{https://doi.org/10.1007/BFb0019850}.
\bibitem{GrasslBZ} M. Grassl, ``Searching for linear codes with large minimum distance,'' in \emph{Discovering Mathematics with Magma}, Algorithms and Computation in Mathematics, vol.~19, pp.~287--313, Springer, 2006. \url{https://doi.org/10.1007/978-3-540-37634-7_13}.
\bibitem{LT} P. Lison\v{e}k and L. Trummer, ``Algorithms for the minimum weight of linear codes,'' \emph{Advances in Mathematics of Communications}, vol.~10, no.~1, pp.~195--207, 2016. \url{https://doi.org/10.3934/amc.2016.10.195}.
\bibitem{ParallelBZ} G. Quintana-Ort\'i, F. Hernando, and F.~D. Igual, ``Algorithm 1033: Parallel implementations for computing the minimum distance of a random linear code on distributed-memory architectures,'' \emph{ACM Transactions on Mathematical Software}, vol.~49, no.~1, pp.~1--24, 2023. \url{https://doi.org/10.1145/3573383}.
\bibitem{QDistRnd} L.~P. Pryadko, V.~A. Shabashov, and V.~K. Kozin, ``QDistRnd: A GAP package for computing the distance of quantum error-correcting codes,'' \emph{Journal of Open Source Software}, vol.~7, no.~71, art.~4120, 2022. \url{https://doi.org/10.21105/joss.04120}.
\bibitem{QuantumBZ} F. Hernando, G. Quintana-Ort\'i, and M. Grassl, ``Fast algorithms and implementations for computing the minimum distance of quantum codes,'' \emph{ACM Transactions on Quantum Computing}, vol.~7, no.~2, art.~12, 2026. \url{https://doi.org/10.1145/3795877}.
\bibitem{DKP} I. Dumer, A.~A. Kovalev, and L.~P. Pryadko, ``Distance verification for classical and quantum LDPC codes,'' \emph{IEEE Transactions on Information Theory}, vol.~63, no.~7, pp.~4675--4686, 2017. \url{https://doi.org/10.1109/TIT.2017.2690381}.
\bibitem{DistanceBench} M. Webster, A. Jacob, and O. Higgott, ``Distance-finding algorithms for quantum codes and circuits,'' \url{https://arxiv.org/abs/2603.22532v2}, 2026.
\bibitem{QEDist} J.~Y. Lee, ``QEDist: Practical exact distance certification for stabilizer codes,'' \url{https://arxiv.org/abs/2611.21709v1}, 2026. \url{https://github.com/yeon-lee/QEDist}, computation commit \texttt{aeb7006147b77}.
\bibitem{Smith} H.~J.~S. Smith, ``On systems of linear indeterminate equations and congruences,'' \emph{Philosophical Transactions of the Royal Society of London}, vol.~151, pp.~293--326, 1861. \url{https://doi.org/10.1098/rstl.1861.0016}.
\bibitem{Seed} K. Okada and K. Kasai, ``APM--PP construction data, $(J,L,P)=(3,8,192)$,'' public construction catalogue, minimal-twist seed attributed to K.~Okada, 2026. \url{https://kasai.ict.eng.isct.ac.jp/pair_partition_cpm_css_codes_20260714.html}; defining input archived in \cite{CodeData}.
\bibitem{CodeData} K. Kasai, ``Increasing Quantum LDPC Code Distance through Explicit Extensions: Code Data,'' version~1.2.0, 20 November 2026. fixed commit \texttt{02401b9135739}: \url{https://github.com/kasaikenta/apm-pp-distance-extensions/tree/02401b91357392409ae32bcc660e325ea3876691}.
\bibitem{EJ} J. Edmonds and E.~L. Johnson, ``Matching, Euler tours and the Chinese postman,'' \emph{Mathematical Programming}, vol.~5, pp.~88--124, 1973. \url{https://doi.org/10.1007/BF01580113}.
\bibitem{CodeEvidence} K. Kasai, ``Exact-distance evidence for explicit quantum LDPC extensions,'' data and audit software, version~1.3.0, 21 November 2026. \url{https://github.com/kasaikenta/apm-pp-distance-extensions/releases/tag/v1.3.0}.
\end{thebibliography}
\end{document}